\documentclass[usegraphicx,usenatbib]{mn2e}

\usepackage[a4paper,total={17.8cm,24.0cm},centering]{geometry}
\usepackage{times}
\usepackage[retainorgcmds]{IEEEtrantools}

\usepackage[colorlinks,citecolor=blue]{hyperref} 
\usepackage[all]{hypcap} 

\newcommand{\dd}{\mathrm{d}}
\newcommand{\apj}{ApJ}           
\newcommand{\apjl}{ApJ}           
\newcommand{\mnras}{MNRAS}       
\newcommand{\aap}{A\&A}
\newcommand{\aj}{AJ}
\newcommand{\araa}{ARA\&A}
\def\equationautorefname~#1\null{Equation~(#1)\null}
\def\sectionautorefname~#1\null{Section~#1\null}

\title[Spherical anisotropic proper motions]{General spherical anisotropic Jeans models of stellar kinematics: including proper motions and radial velocities}

\author[M.~Cappellari]{Michele Cappellari\thanks{E-mail:
michele.cappellari@physics.ox.ac.uk}\\
Sub-Department of Astrophysics, Department of Physics, University of Oxford, Denys Wilkinson Building, Keble Road, Oxford, OX1 3RH}

\date{Posted to arXiv on 1 May 2015}

\pagerange{\pageref{firstpage}--\pageref{lastpage}} \pubyear{2015}

\begin{document}
\label{firstpage}
\maketitle

\begin{abstract}
\citet{Cappellari2008} presented a flexible and efficient method to model the stellar kinematics of anisotropic axisymmetric and spherical stellar systems. The spherical formalism could be used to model the line-of-sight velocity second moments allowing for essentially arbitrary radial variations in the anisotropy and general luminous and total density profiles. Here we generalize the spherical formalism by providing the expressions for all three components of the projected second moments, including the two proper motion components. A reference implementation is now included in the public \textsc{jam} package available at \url{http://purl.org/cappellari/software}.
\end{abstract}


\section{Introduction}

In \citet{Cappellari2008} we used the \citet{Jeans1922} equations to derive the projected second velocity moments for an anisotropic axisymmetric or spherical stellar system for which both the luminous and total densities are described via the Multi-Gaussian Expansion (MGE,  \citealt{Emsellem1994,Cappellari2002mge}). We called the technique the Jeans Anisotropic Modelling (\textsc{JAM}) method and provided a reference software implementation\footnote{Available from \url{http://purl.org/cappellari/software}} (in IDL and Python). An implementation\footnote{Available from \url{https://github.com/lauralwatkins/cjam}} in the C language was provided by \citet{Watkins2013}.

In an addendum \citep{Cappellari2012jam} we gave explicit expression for the 
six projected second moments, including both proper motions and radial velocities, for the axisymmetric case (see also \citealt{DSouza2013,Watkins2013}). All projected components can be written using a single numerical quadrature and without using special functions.

In this short note we do the same for the spherical case, namely we provide explicit expression for the three components of the projected second velocity moments. We adopt identical notation and coordinates system as in \citet{Cappellari2008}, and we refer the reader to that paper for details and definitions. 

\section{Jeans solution with proper motions}

We assume spherical symmetry and constant anisotropy for each individual MGE component. The Jeans equation can then be written \citep[equation~4.215]{Binney2008}
\begin{equation}
\frac{\dd(\nu\overline{v_r^2})}{\dd r}+\frac{2\beta\, \nu\overline{v_r^2}}{r}=-\nu\frac{\dd \Phi}{\dd r},
\end{equation}
where $\overline{v_{\theta}^2}=\overline{v_{\phi}^2}$ for symmetry and we defined $\beta=1-\overline{v_{\theta}^2}/\overline{v_r^2}$.

We use equation~(19) of \citet{vanderMarel2010} which provide the projection expressions for all three components of the velocity second moments, including the proper motions \citep[see also][]{Strigari2007}. We then follow the same steps and definitions as \citet[section~3.2.1]{Cappellari2008} to write all three projected second velocity moments as follows
\begin{eqnarray}
\Sigma \overline{v_{\alpha}^2}(R)=
2G\int_R^{\infty }
\left[\frac{r^{1-2 \beta } \mathcal{Q}_\alpha(r)}{\sqrt{r^2-R^2}}
\int _r^{\infty }\frac{\nu(u)M(u)}{u^{2-2\beta}}\dd u\right] \, \dd r,
\end{eqnarray}
where (i) $\alpha={\rm los}$ for the line-of-sight velocity (ii) $\alpha={\rm pmr}$ for the  radial proper motion, measured from the projected centre of the system, and (iii) $\alpha={\rm pmt}$ for the tangential proper motion, respectively and we defined
\begin{equation}
\mathcal{Q}_{\rm los}(r)=1-\beta\, (R/r)^2
\end{equation}
\begin{equation}
\mathcal{Q}_{\rm pmr}(r)=1-\beta+\beta\, (R/r)^2
\end{equation}
\begin{equation}
\mathcal{Q}_{\rm pmt}(r)=1-\beta.
\end{equation}

Integrating by parts one of the two integrals disappears and all three projected second moments can still be written as in equation~(42) of \citet{Cappellari2008}
\begin{equation}\label{eq:sph_jeans}
\Sigma \overline{v_\alpha^2}(R) = G\int_R^{\infty}\!\!
\mathcal{F}_\alpha\!\! \left(\frac{R^2}{r^2}\right)\nu(r)\,M(r)\, \dd r.
\end{equation}
As shown in \autoref{sec:mge_spherical}, when using the MGE parametrization, the evaluation of this expression requires a single numerical quadrature and some special functions.
For the line-of-sight component the expression for $\mathcal{F}_{\rm los}$ was given by equation~(43) of \citet{Cappellari2008}

\begin{IEEEeqnarray}{rCl}\label{eq:f_los}
\mathcal{F}_{\rm los}(w) & = & \frac{w^{1-\beta }}{R}\left[\beta\, B_w\!\left(\beta+\frac{1}{2} ,\frac{1}{2}\right)-B_w\!\left(\beta-\frac{1}{2} ,\frac{1}{2}\right)\right.\nonumber\\
&& + \left.\frac{\sqrt{\pi }\,(\frac{3}{2}-\beta )\,\Gamma\!\left(\beta-\frac{1}{2} \right)}{\Gamma(\beta )}\right]
\end{IEEEeqnarray}
\citep[see also][]{Mamon2005}, where $\Gamma$ is the Gamma function \citep[equation~6.1.1]{Abramowitz1965} and $B_w$ is the incomplete Beta function \citep[equation~6.6.1]{Abramowitz1965}, for which efficient routines exist in virtually any language. 
The corresponding expressions to use for the radial and tangential proper motion components are
\begin{IEEEeqnarray}{rCl}\label{eq:f_pmr}
\mathcal{F}_{\rm pmr}(w) & = & \frac{w^{1-\beta}}{R}\! \left[(\beta-1) B_w\!\left(\beta-\frac{1}{2} ,\frac{1}{2}\right) - \beta\, B_w\!\left(\beta+\frac{1}{2} ,\frac{1}{2}\right) \right.\nonumber\\
&& + \left.\frac{\sqrt{\pi }\,\Gamma\!\left(\beta - \frac{1}{2} \right)}{2\, \Gamma(\beta )}\right],
\end{IEEEeqnarray}

\begin{equation}\label{eq:f_pmt}
\mathcal{F}_{\rm pmt}(w)=\frac{w^{1-\beta}(\beta-1)}{R}\!\!  \left[B_w\!\left(\beta-\frac{1}{2} ,\frac{1}{2}\right)-\frac{\sqrt{\pi }\,\Gamma\!\left(\beta - \frac{1}{2} \right)}{\Gamma(\beta )}\right]\!\!.
\end{equation}
Specific expressions can be obtained for $\beta=\pm1/2$, where the $B_w$ function is divergent, but in real applications it is sufficient to perturb $\beta$ by an insignificant amount to avoid the singularity.
In the isotropic limit all three components become equal
\begin{equation}
\lim_{\beta\rightarrow0}\mathcal{F}_{\rm los}
=\lim_{\beta\rightarrow0}\mathcal{F}_{\rm pmr}
=\lim_{\beta\rightarrow0}\mathcal{F}_{\rm pmt}
=2\frac{\sqrt{r^2 - R^2}}{r^2}
\end{equation}
and \autoref{eq:sph_jeans} reduces to equation (29) of \citet{Tremaine1994}.

\section{MGE spherical Jeans solution}
\label{sec:mge_spherical}

Here we apply the general spherical Jeans solution with constant anisotropy to derive an actual solution for a stellar system in which both the luminous density and the total one are described by the MGE parametrization. In this case the surface brightness $\Sigma_k$, the luminosity density $\nu_k$ and the total density $\rho_j$ for each individual Gaussian are given by \citep{Bendinelli1991}
\begin{equation}\label{eq:surf_sph}
    \Sigma_k(R) =
    \frac{L_k}{2\pi\sigma^2_k}
    \exp\left(-\frac{R^2}{2\sigma^2_k}\right),
\end{equation}
\begin{equation}\label{eq:dens_sph}
    \nu_k(r) =
    \frac{L_k}{(\sqrt{2\pi}\, \sigma_k)^3}
    \exp\left(-\frac{r^2}{2\sigma_k^2}\right),
\end{equation}
\begin{equation}\label{eq:mass_sph}
    \rho_j(r) =
    \frac{M_j}{(\sqrt{2\pi}\, \sigma_j)^3}
    \exp\left(-\frac{r^2}{2\sigma_j^2}\right).
\end{equation}
The mass of a Gaussian contained within the spherical radius $r$ is given by equation~(49) of \citet{Cappellari2008}
\begin{equation}\label{eq:massr}
    M_j(r)=M_j \left[{\rm erf}\left(\frac{r}{\sqrt{2}\, \sigma_j}\right)
    -\frac{\sqrt{\frac{2}{\pi}}\; r}{\sigma_j}
    \exp\left(-\frac{r^2}{2 \sigma_j^2}\right)
    \right],
\end{equation}
with ${\rm erf}(x)$ the error function \citep[equation~7.1.1]{Abramowitz1965}.

The projected second velocity moments for the whole MGE model, summed over all the $N$ luminous and $M$ massive Gaussians, for any of the three velocity second moment components, are still given by equation~(50) of \citet{Cappellari2008}
\begin{equation}\label{eq:sph_jeans_mge}
\Sigma \overline{v_\alpha^2}(R) = G\!\!\int_R^{\infty}\!\!
\sum_{k=1}^N\! \mathcal{F}_{\alpha,k}\!\!\left(\frac{R^2}{r^2}\right)\!\nu_k(r)\!\!
\left[M_\bullet\!+\!\sum_{j=1}^M\! M_j(r)\right]\!
\dd r,
\end{equation}
where $\nu_k(r)$ is given by \autoref{eq:dens_sph}, $M_j(r)$ is given by \autoref{eq:massr}, and $\mathcal{F}_{\alpha,k}$  is obtained by replacing the $\beta$ parameter in \autoref{eq:f_los}--(\ref{eq:f_pmt}) with the anisotropy $\beta_k$ of each luminous Gaussian component of the MGE.

The formalism presented in this section was implemented in an updated version of the public \textsc{jam} package (see footnote~1).

\section{Application}

\begin{figure}
\includegraphics[width=\columnwidth]{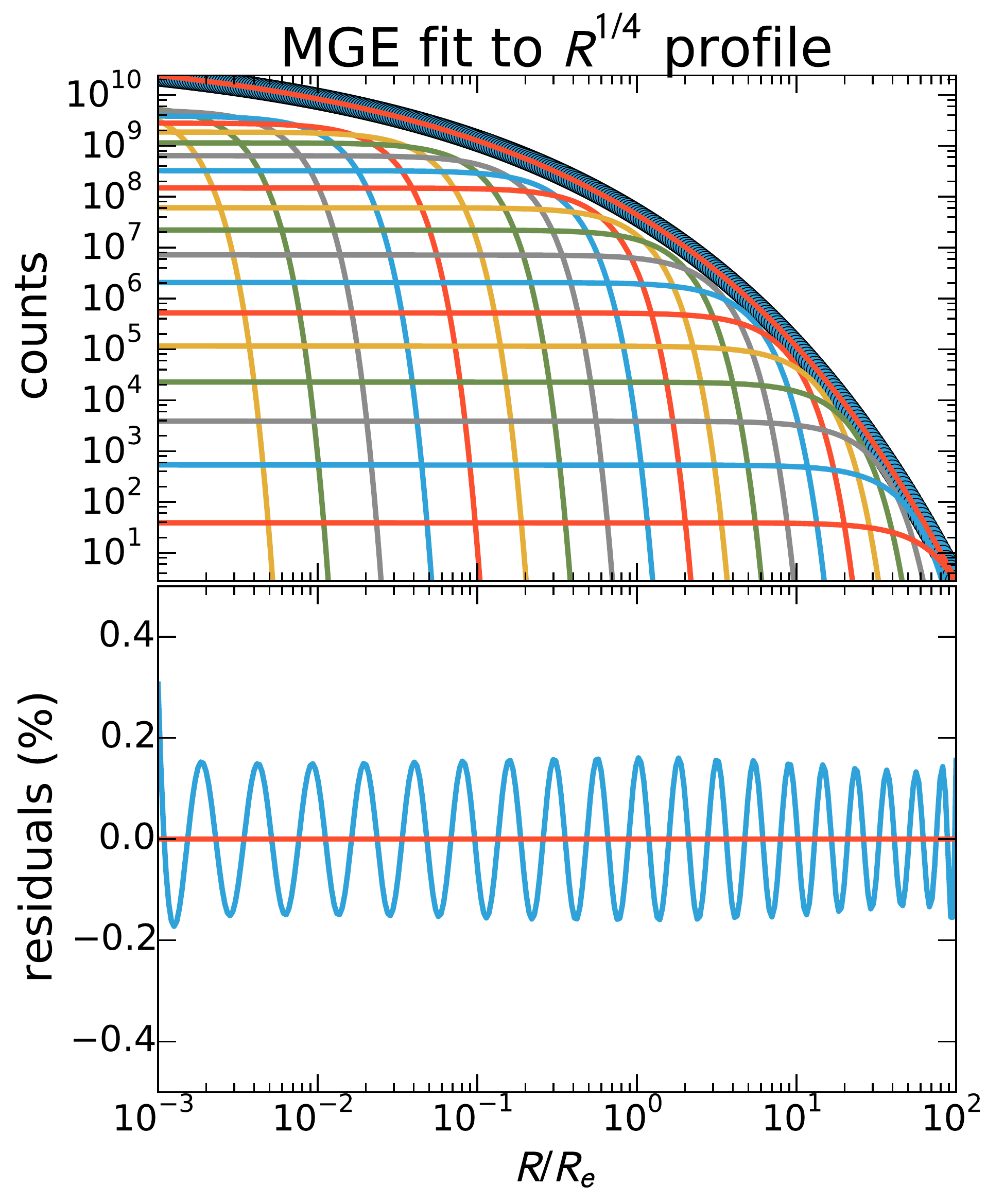}
\caption{MGE fit to a \citet{deVaucouleurs1948} $R^{1/4}$ profile. We used 20 Gaussians and the \textsc{mge\_fit\_1d} routine to describe the profile in the range $-3<\log R/R_{\rm e}<2$ with a maximum relative error of about 0.2\%. The resulting MGE model contains 99.998\% of the analytic total mass.}
\label{fig:mge}
\end{figure}

\begin{figure}
\includegraphics[width=\columnwidth]{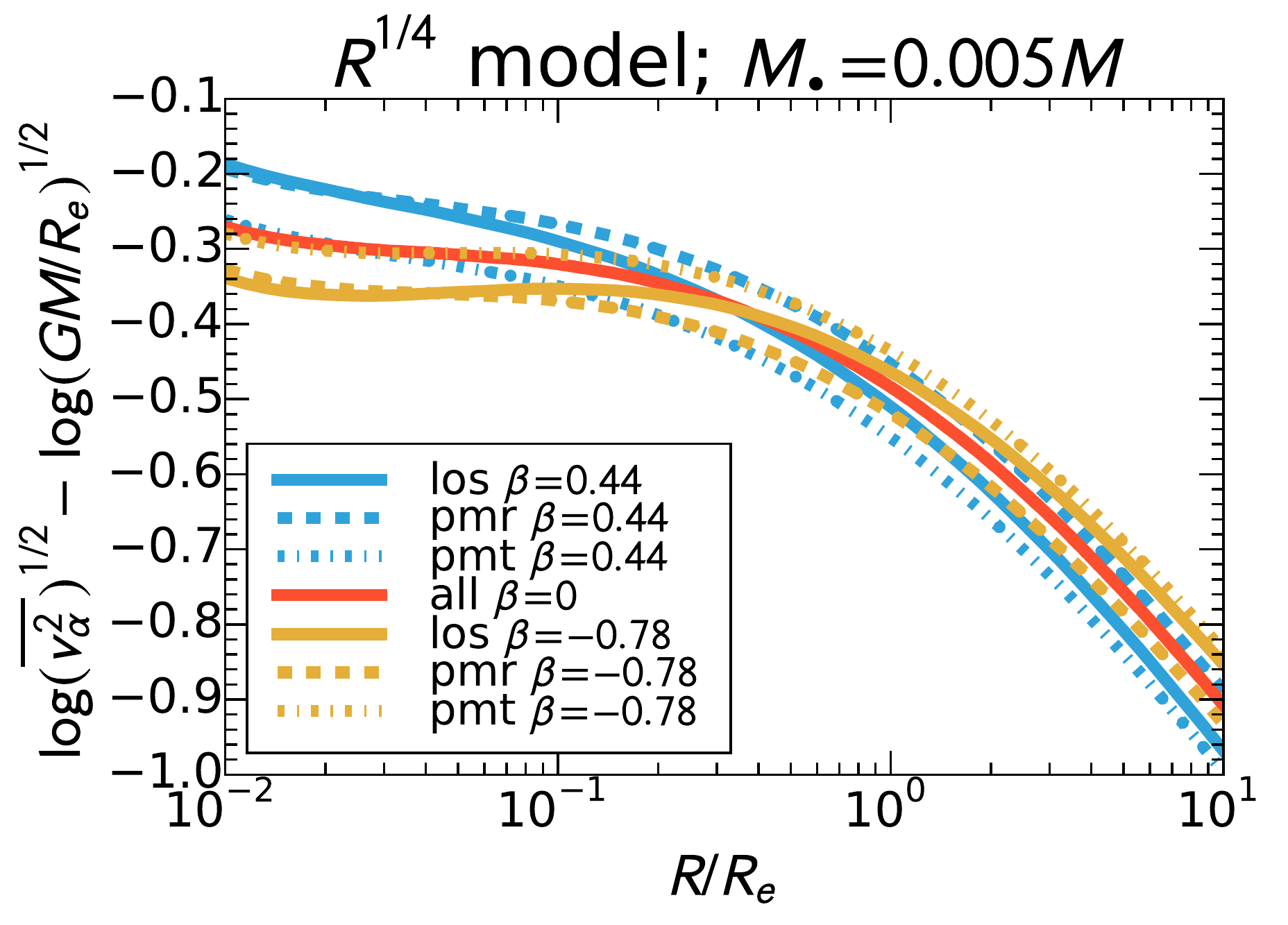}
\includegraphics[width=\columnwidth]{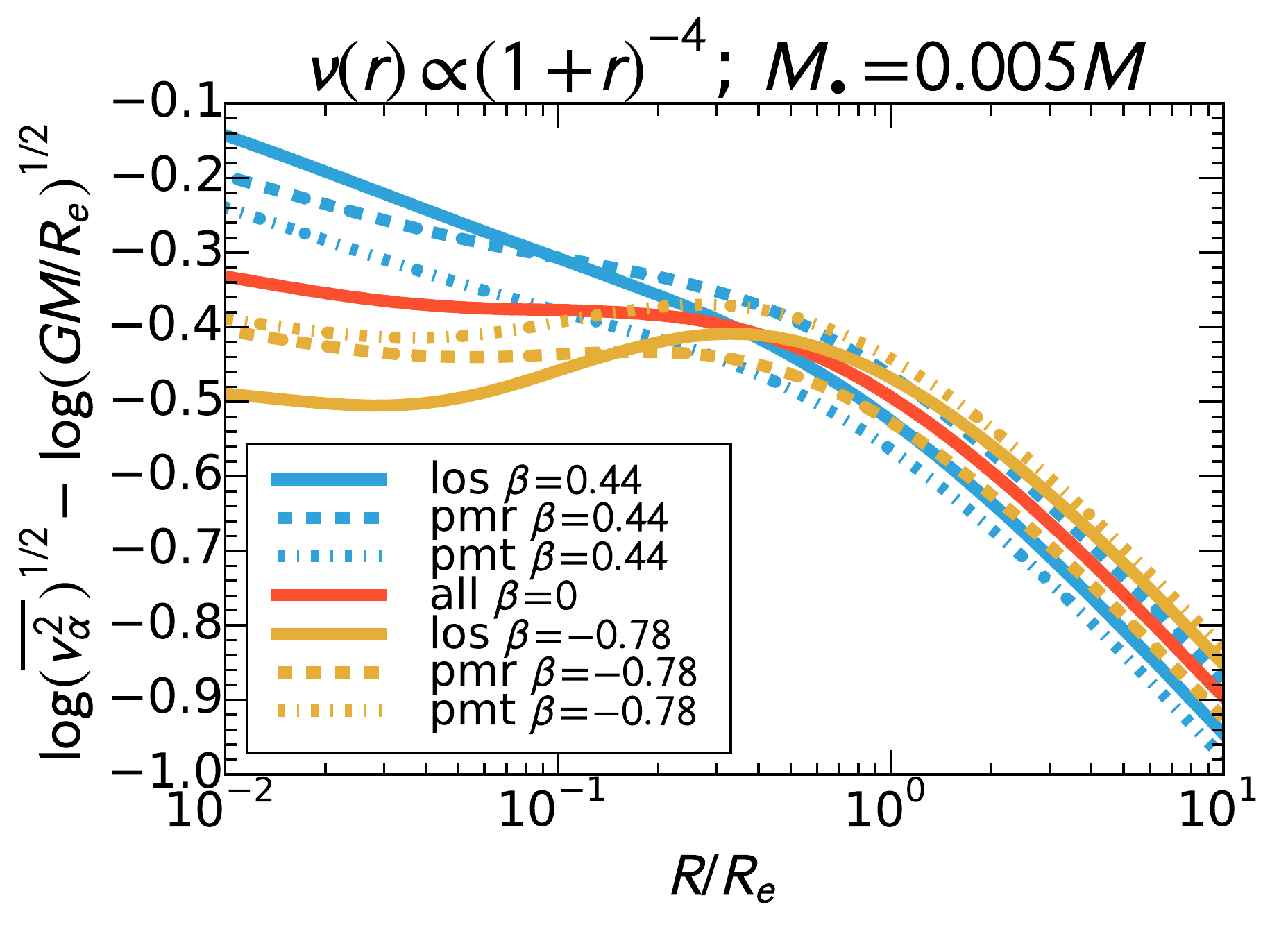}
\caption{Projected velocity second moments $\overline{v_\alpha^2}(R)$ as a function of radius, normalized by the half-light radius $R_{\rm e}$, for a range of anisotropies $\beta$, for spherical models with a central black hole mass $M_\bullet$ of 0.5\% of the total mass. The {\em top panel} is for a \citet{deVaucouleurs1948} $R^{1/4}$ surface brightness profile, while the {\em bottom panel} is for a `cored' luminosity density profile, with asymptotically constant nuclear density. The three adopted anisotropies $\beta$ were chosen to correspond to axial ratios of the velocity ellipsoid of $(\overline{v_{\theta}^2}/\overline{v_r^2})^{1/2}=3/4$, 1 and $4/3$ respectively. Here $\alpha={\rm los}$, pmr or pmt for the line-of-sight, the radial or tangential proper motions respectively. The presence of a core {\em enhances} the observable effects of $\beta$ variations and {\em reduces} the ability of the proper motions to measure $\beta$, especially for $\beta<0$. This makes the determination of mass profiles more challenging in cored than in cusped luminosity profiles.}
\label{fig:jam}
\end{figure}

As an illustration of the general behaviour of the second moments, we used the Python version of the \textsc{mge\_fit\_1d} one-dimensional fitting routine in the \textsc{mge\_fit\_sectors} package \citep[see footnote~1]{Cappellari2002mge} to obtain an accurate MGE description of a \citet{deVaucouleurs1948} $R^{1/4}$ profile (\autoref{fig:mge}). We then used the spherical \textsc{jam\_sph\_rms} routine in the \textsc{jam} package (see footnote~1) to calculate the predicted velocity second moments for different (here constant) anisotropies, using \autoref{eq:sph_jeans_mge}. The model assumes that mass follows light except for the presence of a central black hole with mass $M_\bullet=0.005\times M$, where $M$ is the total mass of the system. The fractional value is the typical observed one from \citet{Kormendy2013review}. The resulting predicted second velocity moment profiles are shown in the top panel of \autoref{fig:jam}. The bottom panel is like the top one, for a `cored' profile $\nu(r)\propto(1+r)^{-4}$ with asymptotically constant luminosity density at small radii.

For the line-of-sight component, \autoref{fig:jam} illustrates the well known mass-anisotropy degeneracy: the profile changes significantly at fixed $M_\bullet$. If the anisotropy is unknown, $M_\bullet$ cannot be measured from these spherical models. However, using proper motions one can constrain the anisotropy directly and consequently measure $M_\bullet$ using the second velocity moments alone \citep{vanderMarel2010}. The bottom panel also shows that models are more challenging for more shallow inner profiles due to the reduced sensitivity of the ratio $(\overline{v_{\rm pmt}^2}/\overline{v_{\rm pmr}^2})^{1/2}$ to $\beta$ variations.

In \autoref{fig:jam} all three components are plotted in the same units as 
for the radial velocity. In a real application of the method, the velocities of 
the proper motion components need to be converted into proper motions units, 
using the distance of the system. The distance is a free model parameter and this allows one to measure kinematical distances \citep[e.g.][]{vandeVen2006,vandenBosch2006,vanderMarel2010}.

By assigning different anisotropies $\beta_k$ to the different Gaussians one can describe essentially arbitrary smooth radial variations of the anisotropy  \citep[e.g.][section~3.3]{Cappellari2009cena}. While, by using different Gaussians for the luminous and total density, one can describe general dark halo profiles \citep[e.g.][]{Cappellari2015dm}. In practice, for spherical geometry, one can perform a high-accuracy MGE fit to some parametric description of the dark and luminous profiles, using the \textsc{mge\_fit\_1d} fitting program of \citet[see footnote~1]{Cappellari2002mge} as shown in \autoref{fig:mge}.

\section{Conclusions}

Proper motion data have the fundamental advantage over the line-of-sight quantities that they allow one to break, using the velocity second moments alone, the mass-anisotropy degeneracy affecting the recovery of mass profiles in spherical systems \citep{Binney1982}.
The proposed formalism is becoming useful to model the growing number of proper motion data becoming available. These are now being provided mainly by the Hubble Space Telescope \citep[e.g.][]{Watkins2015}. In the near future proper motion data for the Milky Way satellites will also be provided by the GAIA spacecraft, while in the more distant future a major step forward in proper motion determinations will be made by EUCLID space mission. 

\section*{Acknowledgements}

I acknowledge support from a Royal Society University Research Fellowship. This paper made use of matplotlib \citep{matplotlib2007}.

\label{lastpage}


\begin{thebibliography}{23}
\expandafter\ifx\csname natexlab\endcsname\relax\def\natexlab#1{#1}\fi

\bibitem[{{Abramowitz} \& {Stegun}(1964){Abramowitz} M., {Stegun}
  I.~A.}]{Abramowitz1965}
{Abramowitz} M., {Stegun} I.~A., 1964, Handbook of mathematical functions
  (Reprinted 1972), Dover Books on Advanced Mathematics. Dover, New York

\bibitem[{{Bendinelli}(1991){Bendinelli} O.}]{Bendinelli1991}
{Bendinelli} O., 1991, \apj, 366, 599

\bibitem[{{Binney} \& {Mamon}(1982){Binney} J., {Mamon} G.~A.}]{Binney1982}
{Binney} J., {Mamon} G.~A., 1982, \mnras, 200, 361

\bibitem[{{Binney} \& {Tremaine}(2008){Binney} J., {Tremaine} S.}]{Binney2008}
{Binney} J., {Tremaine} S., 2008, Galactic Dynamics: Second Edition. Princeton
  University Press, Princeton, NJ

\bibitem[{{Cappellari}(2002){Cappellari} M.}]{Cappellari2002mge}
{Cappellari} M., 2002, \mnras, 333, 400

\bibitem[{{Cappellari}(2008){Cappellari} M.}]{Cappellari2008}
{Cappellari} M., 2008, \mnras, 390, 71

\bibitem[{{Cappellari}(2012){Cappellari} M.}]{Cappellari2012jam}
{Cappellari} M., 2012, e-print (arXiv:1211.7009)

\bibitem[{{Cappellari} {et~al}\mbox{.}(2009){Cappellari} M.
  {et~al.}}]{Cappellari2009cena}
{Cappellari} M., {Neumayer} N., {Reunanen} J., {van der Werf} P.~P., {de Zeeuw}
  P.~T., {Rix} H.-W., 2009, \mnras, 394, 660

\bibitem[{{Cappellari} {et~al}\mbox{.}(2015){Cappellari} M.
  {et~al.}}]{Cappellari2015dm}
{Cappellari} M. {et~al.}, 2015, ApJL, in press (arXiv:1504.00075)

\bibitem[{{de Vaucouleurs}(1948){de Vaucouleurs} G.}]{deVaucouleurs1948}
{de Vaucouleurs} G., 1948, Annales d'Astrophysique, 11, 247

\bibitem[{{D'Souza} \& {Rix}(2013){D'Souza} R., {Rix} H.-W.}]{DSouza2013}
{D'Souza} R., {Rix} H.-W., 2013, \mnras, 429, 1887

\bibitem[{{Emsellem}, {Monnet} \& {Bacon}(1994){Emsellem} E., {Monnet} G.,
  {Bacon} R.}]{Emsellem1994}
{Emsellem} E., {Monnet} G., {Bacon} R., 1994, \aap, 285, 723

\bibitem[{Hunter(2007)Hunter J.~D.}]{matplotlib2007}
Hunter J.~D., 2007, Computing In Science \& Engineering, 9, 90

\bibitem[{{Jeans}(1922){Jeans} J.~H.}]{Jeans1922}
{Jeans} J.~H., 1922, \mnras, 82, 122

\bibitem[{{Kormendy} \& {Ho}(2013){Kormendy} J., {Ho}
  L.~C.}]{Kormendy2013review}
{Kormendy} J., {Ho} L.~C., 2013, \araa, 51, 511

\bibitem[{{Mamon} \& {{\L}okas}(2005){Mamon} G.~A., {{\L}okas}
  E.~L.}]{Mamon2005}
{Mamon} G.~A., {{\L}okas} E.~L., 2005, \mnras, 363, 705

\bibitem[{{Strigari}, {Bullock} \& {Kaplinghat}(2007){Strigari} L.~E.,
  {Bullock} J.~S., {Kaplinghat} M.}]{Strigari2007}
{Strigari} L.~E., {Bullock} J.~S., {Kaplinghat} M., 2007, \apjl, 657, L1

\bibitem[{{Tremaine} {et~al}\mbox{.}(1994){Tremaine} S.
  {et~al.}}]{Tremaine1994}
{Tremaine} S., {Richstone} D.~O., {Byun} Y.-I., {Dressler} A., {Faber} S.~M.,
  {Grillmair} C., {Kormendy} J., {Lauer} T.~R., 1994, \aj, 107, 634

\bibitem[{{van de Ven} {et~al}\mbox{.}(2006){van de Ven} G.
  {et~al.}}]{vandeVen2006}
{van de Ven} G., {van den Bosch} R.~C.~E., {Verolme} E.~K., {de Zeeuw} P.~T.,
  2006, \aap, 445, 513

\bibitem[{{van den Bosch} {et~al}\mbox{.}(2006){van den Bosch} R.
  {et~al.}}]{vandenBosch2006}
{van den Bosch} R., {de Zeeuw} T., {Gebhardt} K., {Noyola} E., {van de Ven} G.,
  2006, \apj, 641, 852

\bibitem[{{van der Marel} \& {Anderson}(2010){van der Marel} R.~P., {Anderson}
  J.}]{vanderMarel2010}
{van der Marel} R.~P., {Anderson} J., 2010, \apj, 710, 1063

\bibitem[{{Watkins} {et~al}\mbox{.}(2013){Watkins} L.~L.
  {et~al.}}]{Watkins2013}
{Watkins} L.~L., {van de Ven} G., {den Brok} M., {van den Bosch} R.~C.~E.,
  2013, \mnras, 436, 2598

\bibitem[{{Watkins} {et~al}\mbox{.}(2015){Watkins} L.~L.
  {et~al.}}]{Watkins2015}
{Watkins} L.~L., {van der Marel} R.~P., {Bellini} A., {Anderson} J., 2015,
  \apj, 803, 29

\end{thebibliography}
\end{document}